\documentclass[aps,prb,twocolumn,groupedaddress,showkeys,floatfix]{revtex4}
\usepackage{graphicx}

\begin{document}


\title{Anisotropic coarse-grained statistical potentials improve the ability to identify native-like protein structures}

\author{N.-V. Buchete}
\affiliation{Department of Chemistry, Boston University, Boston, MA 02215}
\author{J.E. Straub}
\affiliation{Department of Chemistry, Boston University, Boston, MA 02215}
\author{D. Thirumalai}
\affiliation{Institute for Physical Science and Technology, University of Maryland,
College Park, MD 20742}

\date{\today}

\begin{abstract}
We present a new method to extract distance and orientation dependent potentials between amino acid side chains using a database of protein structures and the standard Boltzmann device. The importance of orientation dependent interactions is first established by computing orientational order parameters for proteins with $\alpha$-helical and $\beta$-sheet architecture. Extraction of the anisotropic interactions requires defining local reference frames for each amino acid that uniquely determine the coordinates of the neighboring residues. Using the local reference frames and histograms of the radial and angular correlation functions for a standard set of non-homologue protein structures, we construct the anisotropic pair potentials. The performance of the orientation dependent potentials was studied using a large database of decoy proteins. The results demonstrate that the new distance and orientation dependent residue-residue potentials present a significantly improved ability to recognize native folds from a set of native and decoy protein structures.\end{abstract}



\maketitle


\section{Introduction}

	A major goal of proteomics is to determine structures of proteins rapidly. It is impractical to obtain very high resolution structures on a genome wide scale. Furthermore, for processing biological functions it is important to characterize the dynamics and thermodynamics of a network of proteins. From a computational perspective, it is currently impossible to realize these goals using all-atom molecular dynamics simulations \cite{Kollman-98,Ferrara}. Thus, there is an urgent need to develop coarse-grained, yet reliable, models for protein structures.  Construction of such models requires the determination of interaction potentials between amino acid residues. The wealth of structural data on a number of proteins in the Protein Data Bank (PDB) \cite{PDB} has been a source for obtaining interaction potentials \cite{Scheraga-01,Jernigan-99,Skolnick-01}.

	The idea of using the frequencies of  amino acid pairing to determine potential interaction parameters was first proposed by Tanaka and Scheraga \cite{Scheraga76}. With the exception of a few studies \cite{Sippl90}, most of the ``knowledge-based'' potentials have been obtained  solely in terms of residue-residue contacts. 
 Sippl \cite{Sippl90} and others have introduced an explicit distance dependence in the database-derived mean force potentials using the Boltzmann formula. This method, known as the ``Boltzmann device'', assumes that the known protein structures from the PDB correspond to classical equilibrium states. The side chain - side chain (SC-SC) potentials can be related to distance-dependent probability densities $f(r)$ by the relation
\begin{equation}
U^{ij}(r)=-kT\ ln\left[\frac{f^{ij}(r)}{f_{ref}(r)}\right]
\label{eq:boltzmann}
\end{equation}
where $f^{ij}(r)$ is the probability density for a side chain of type $i$ to be separated by a distance $r$ from a side chain of type $j$, and the choice of the reference probability density $f_{ref}(r)$ is very important \cite{Thirum-BT99,Sippl95}. Sippl \cite{Sippl95} suggests that $r$ can be the distance between two atoms or some other structural parameter such as dihedral angles. Other statistical potentials developed subsequently using this approach have interpreted $f(r)$  as the distance-dependent probability density. 

	In recent years, a number of studies have evaluated the goodness of statistical potentials based on pairwise additive interactions between residues \cite{Thirum-BT99,Vendruscolo,Ben-Naim,Elber-00b,Elber-00c,Elber-01a,Elber-01b,Vajda-00}.
These studies have concluded that pairwise additive potentials, dependent only on the radial distance between residues, are inadequate for structure prediction. One of the major drawbacks of using contact potentials or potentials that only depend on the radial distance is that the relative orientation between the amino acids, which plays a role in the packing of the protein interior, is not taken into account. To account for the dense interior of the folded states of proteins it is crucial to optimize the relative orientation of the participating side chains \cite{Richards-94}. 
It is the purpose of this paper to examine the relevance of interaction potentials between amino acids that include the orientational dependence.

	By analyzing the angular distribution of side chains around amino acids, Bahar and Jernigan \cite{Jernigan-96} showed that some residue pairs have specific coordination states with much higher probabilities than those expected from random distributions. Another indication that the relative residue-residue orientations are important is the recent success of the united-residue force field (UNRES) developed by Scheraga and coworkers \cite{Unres-1,Unres-2,Unres-3}. The SC-SC interactions of the UNRES potentials are parameterized as van der Waals potentials with a pair-specific angular dependence included in the well depth. The UNRES potentials employ a generalized Gay-Berne potential \cite{GayBerne-1,Vorobjev-1,Vorobjev-2} that assumes that the interacting sites are  ellipsoids of revolution placed at the center of mass of the side chains. Although the relative side chain orientations are described in a simplified manner, the success  of the UNRES potentials demonstrates the importance of including orientational dependence in the side chain interaction potential. 

	This idea is also supported by recent studies of protein side chain packing \cite{Shakhnovich-01} which found that current potentials can accommodate more than one rotamer for 95\% of side chain positions. These studies suggest that interaction potentials that can discriminate between a large number of competing rotamer states are required.

	The main goals of this paper are (a) to extract statistical information about  the relative residue-residue orientations and distances in proteins using the available high resolution PDB structures and (b) to investigate the utility of the orientational information in enhancing the ability of distance dependent potentials to identify native protein folds. These goals can contribute to a better understanding of the specific residue packing of native protein structures. Motivated by the above mentioned results of Jernigan et al. \cite{Jernigan-96} and Scheraga et al. \cite{Unres-1}, we approached the first goal by defining local reference frames (LRFs) that permit a more precise, quantitative description of the relative orientation of a given pair of side chains. The new LRFs are related more closely to the three-dimensional configuration of the individual side chains, rather than to their relative position with respect to the backbone or to the neighboring residues. We could extract thus novel radial and angular dependent inter-residue statistical potentials that reflect the specific  distributions of distances and relative SC-SC orientations as observed in native structures of real proteins. In order to address our second goal we tested the efficacy of the new potentials by using a large database of incorrect models of real proteins \cite{Hendlich,Samudrala}, known as decoys. These decoys are computer-generated models of protein structures, specifically designed for being used in evaluating the capacity of various potential functions to distinguish the native-like conformations from non-native ones.

	Before presenting our methods and the results that we obtained, a few clarifications are necessary. First, a complete set of potentials for coarse-grained protein folding simulations should take into account interactions between side chains, interactions between side chains and peptide groups \cite{Keskin98F&D}, and energetics dependent on torsional angles that define the backbone structure \cite{Unres-1,Unres-2,Unres-3}. We only investigate the specific features of SC-SC interactions, as has been done before in studies of statistical potentials that are only contact or distance dependent.
Secondly, we constructed our own corresponding set of distance-dependent potentials rather than employing potentials constructed by other authors. In this manner we minimized the possibility that a better parameter fitting, specific computational implementation, or other technical aspects could affect our results. Finally, as in similar studies \cite{Dill-96} we do not directly address the issue of the dependence of the results on the training database size or sampling problems. Instead, we used a standard, reproducible approach, by employing the set of non-homologous proteins that was used by Scheraga et al. \cite{Unres-1,Unres-2,Unres-3} for similar purposes.

\section{Methods}

	The relative residue-residue orientations are thought to be directly related to the nature of the forces that shape the specific three-dimensional structures of proteins \cite{Jernigan-96,Unres-1}. However, a quantitative approach to the statistical extraction of these orientational information from high resolution structures is still needed. The necessity of including orientation dependent interactions is established in section \ref{sec:OOPs} by computing correlation functions that probe orientational order in the PDB structures. It is shown that, for a given native state topology, the orientational packing of side chains may be decomposed into a linear combination of simple cluster geometries. However, the use of simple orientational order parameters is \emph{not sufficient} for discriminating between basic protein architectures such as $\alpha$-helix or $\beta$-sheet. As such, we are motivated to use more detailed quantitative descriptions of relative residue orientations. In order to achieve this goal, in section \ref{sec:LRFs} we introduce definitions of amino acid dependent local reference frames (LRFs) that permit a standard description of the relative SC-SC orientations. A method for extracting the radial and angular dependent pair distribution functions is presented in section \ref{sec:Pot}, emphasizing the limits imposed on the statistical analysis by the accuracy of the available experimental database of protein structures.

\subsection{Measures of Orientational Order. The Dependence of Anisotropy on Native State Topology}
\label{sec:OOPs}

	The specificity of SC-SC contacts suggests that the relative residue orientations should play a significant role in determining packing in proteins. This idea, supported by the study of Bahar and Jernigan \cite{Jernigan-96}, builds up on the more general theme of how side chains pack in the native states of proteins \cite{Richards-94,Shakhnovich-01}. 
The observation that the interior of protein structures is densely packed \cite{Soyer-00,Levitt-97,Tsai-99} raises the question if one can use a simple crystal lattice description (e.g. $sc$, $bcc$, etc.) for the side chain packing.
While there are many qualitative descriptions of residue packing, they are mainly based on the notion that hydrophobic interactions are responsible for globular shapes \cite{Dill-96} and that specific polar interactions and hydrogen bonding are important stabilizing factors. Because these studies do not address quantitatively the issue of relative side chain orientations, we used orientational order parameters (OOPs) \cite{Steinhardt-83} to asses their importance.
 

The Orientational Order Parameters (OOPs) were introduced \cite{Steinhardt-83} to analyze the internal structure of liquids and glasses and have been used as reaction coordinates for studying the structure of a nucleating system \cite{Frenkel-92, Frenkel-95, Frenkel-96}. The OOPs are defined in terms of ``bonds'' that connect a central particle to its neighbors. For each ``bond'', one can compute the corresponding values of the spherical harmonic functions $Y_{lm}(\theta,\phi)$ that can be used as \emph{local order} parameters
\begin{equation}
Q_{lm}({\bf r})\equiv Y_{lm}(\theta({\bf r}),\phi({\bf r})) \mbox{\ .}
\end{equation}
where ${\bf r}$ is the position vector of the central particle and $\theta(r)$ and $\phi(r)$ are the polar and azimuthal angles of vector ${\bf r}$ with respect to any reference frame (see, e.g. \cite{Frenkel-92}). To make the $Q_{lm}$ values representative for describing the orientational order of an entire system of particles, the \emph{global} orientational parameters are defined as
\begin{equation}
\bar{Q}_{lm}\equiv \frac{1}{N_b}\sum_{i=1}^{N_b}Q_{lm}({\bf r}_i)
\end{equation}
where $N_b$ is the total number of bonds in the system.

	Both $Q_{lm}$ and $\bar{Q}_{lm}$ depend on the choice of local reference frame in which the $Y_{lm}$ values are calculated. Second- and third-order rotationally invariant combinations can be constructed using
\begin{equation}
Q_l\equiv \left(\frac{4\pi}{2l+1}\sum_{m=-l}^{l}\left|\bar{Q}_{lm}\right|^2\right)^{1/2}
\end{equation}
and
\begin{equation} 
\begin{array}{l}
W_l \equiv \nonumber \\ 
\sum_{\stackrel{\scriptstyle{m_1,m_2,m_3}}{m_1+m_2+m_3=0}}
\left(\begin{array}{ccc}l&l&l\\m_1&m_2&m_3\end{array}\right)
\cdot\bar{Q}_{lm_1}\bar{Q}_{lm_2}\bar{Q}_{lm_3}
\end{array}
\end{equation}
where the term $\left(\begin{array}{ccc}l&l&l\\m_1&m_2&m_3\end{array}\right)$ is the Wigner-3j symbol \cite{Steinhardt-83}.
In liquids, the ratio
\begin{equation}
\hat{W}_l\equiv W_l / \left(\sum_{m=-l}^{l}\left|\bar{Q}_{lm}\right|^2\right)^{3/2}
\end{equation}
is not sensitive to the exact definition of neighboring bonds of a particle.
The orientational order parameters $Q_l$ and $W_l$, together with the reduced order parameter $\hat{W}_l$ have specific values for a number of simple cluster geometries such as face-centered-cubic (fcc), hexagonal close-packed (hcp), body-centered-cubic (bcc), simple cubic (sc) and icosahedral (icos).
If only the spherical term ($l=0$) dominates, then it is clear that orientational potentials are not expected to be important. The emergence of non-zero values of $Q_l$ and $W_l$ not only signify that side chain orientations are important in dense packing but may also point to the nature of orientational ordering.  

\begin{figure}
\includegraphics{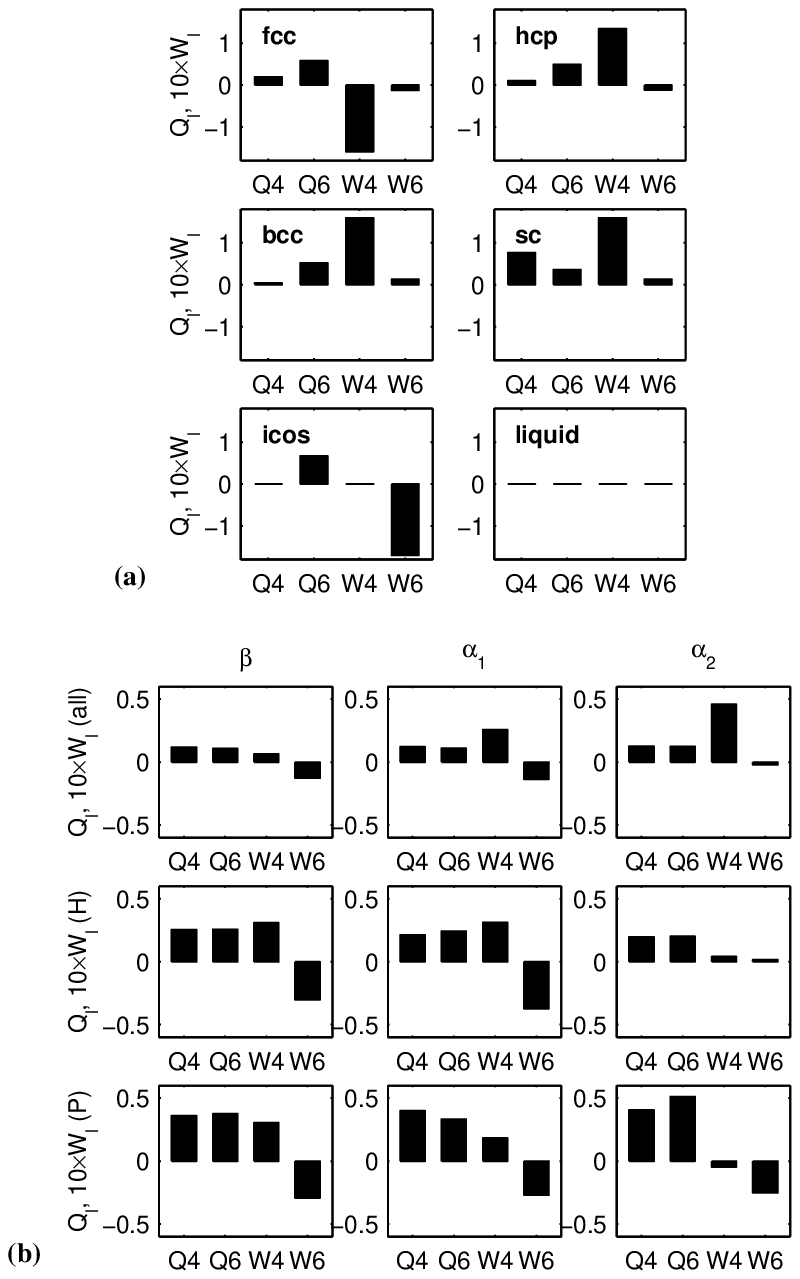}
\caption{\label{fig:QLMs}
{{\bf (a)} The values of the $Q_4$, $Q_6$, $\hat{W}_4$ and $\hat{W}_6$ orientational order parameters for simple cluster geometries \cite{Frenkel-96}. Their specific differences permit a quantitative analysis of the internal order in atomic and molecular systems. {\bf (b)} The values of the $Q_4$, $Q_6$, $\hat{W}_4$ and $\hat{W}_6$ orientational order parameters estimated for three different sets of proteins with $\alpha$-helical and $\beta$-strand architectures (columns). The three rows correspond to averages calculated for all the types of side chains (top), for the five most hydrophobic side chains only (middle), and for the five most hydrophilic residues (bottom). 
The values of $\hat{W}_4$ and $\hat{W}_6$ were multiplied by 10.}}
\end{figure}

All the low-order values ($l \leq 10$) of the parameters $Q_l$, $W_l$ and $\hat{W}_l$ are sensitive to the structural details that are specific to the simple cluster geometries mentioned above \cite{Steinhardt-83,Frenkel-92}. However, for practical reasons, monitoring the values of $Q_4$, $Q_6$, $\hat{W}_4$ and $\hat{W}_6$ suffices to discriminate between the different cluster geometries \cite{Frenkel-95,Frenkel-96}. The values of these four OOPs (Fig. \ref{fig:QLMs}a) serve as a reference to which the values computed using PDB structures can be compared.
Due to scale differences between $Q_l$ and $\hat{W}_l$, the values of $\hat{W}_4$ and $\hat{W}_6$ were multiplied by 10. 

	We calculated the $Q_4$, $Q_6$, $\hat{W}_4$ and $\hat{W}_6$ parameters (shown in Fig. \ref{fig:QLMs}b) for three sets of 50 proteins from the same family with $\alpha$-helical and $\beta$-strand architectures listed in Table \ref{tab:table1}. For each set, we have specifically selected homologue structures to maximize the possibility of finding OOPs that have different, characteristic values for different protein architectures. 
\begin{table}
\caption{\label{tab:table1} The three sets of protein structures used to investigate if a simple packing description is appropriate for protein residues. Highly homologue sequences corresponding to alpha and beta architectures were specifically selected.}
\begin{ruledtabular}
\begin{tabular}{cccccc}
\multicolumn{2}{c}{Set 1 - Ig($\beta$)}&
\multicolumn{2}{c}{Set 2 - Hb($\alpha1$)}&
\multicolumn{2}{c}{Set 3 - Mb($\alpha2$)}\\
\hline
12E8&1CFV&1A00&1CG5&101M&1CH1\\
1A2Y&1CLO&1A01&1CG8&102M&1CH2\\
1A3L&1DCL&1A0U&1CLS&103M&1CH3\\
1A3R&1DSF&1A0V&1DSH&104M&1CH5\\
1A4J&1DVF&1A0W&1DXT&105M&1CH7\\
1A6V&1EUR&1A0X&1DXU&106M&1CH9\\
1A6W&1EUS&1A0Y&1DXV&107M&1CIK\\
1A7N&1F58&1A0Z&1ECA&109M&1CIO\\
1A7O&1FGV&1A3N&1ECD&110M&1CO8\\
1A7P&1FLR&1A3O&1ECN&111M&1CO9\\
1A7Q&1FLT&1A4F&1ECO&112M&1CP0\\
1A7R&1FVC&1AJ9&1FLP&1A6G&1CP5\\
1AJ7&1GIG&1ASH&1FSL&1A6K&1CPW\\
1AP2&1GPO&1AXF&1GBU&1A6M&1EMY\\
1AQK&1HCV&1B2V&1GBV&1A6N&1FCS\\
1AXT&1HIL&1BAB&1GDI&1ABS&1HJT\\
1AY1&1HYX&1BBB&1GDJ&1AJG&1HRM\\
1B0W&1HYY&1BIJ&1GDK&1AJH&1HSY\\
1BFV&1IAM&1BIN&1GDL&1AZI&1IOP\\
1BJM&1IGD&1BUW&1HAB&1BJE&1IRC\\
1BM3&1IGM&1BZ0&1HBA&1BVC&1JDO\\
1BRE&1IND&1BZ1&1HBB&1BVD&1LHS\\
1BWW&1IVL&1BZZ&1HBG&1BZ6&1LHT\\
1CDY&1KEL&1CBL&1HBH&1BZP&1LTW\\
1CFB&1KEM&1CBM&1HBI&1BZR&1M6C\\
\end{tabular}	
\end{ruledtabular}
\end{table}
If similar OOP values were obtained for the homologue sets of hemoglobins (Hb) and myoglobins (Mb), yet different from the ones measured for immunoglobulins (Ig), it would be a strong indication that the OOPs are a sensitive measure of protein architecture. As shown next, this is not the case and, therefore, the more detailed investigation of the relative orientations of side chains is necessary.
The three rows in Fig. \ref{fig:QLMs}b correspond to averages calculated: (a) for all the types of side chains (top), (b) only for five highly hydrophobic side chains
(middle), and (c) for five hydrophilic residues (bottom). The hydrophobic
side chains considered were Ile, Leu, Val, Phe and Met, while the polar
residues were Asn, Gln, Ser, Thr and His. In Fig. \ref{fig:QLMs}b, the
columns correspond to the three sets of proteins that were analyzed. The
first set had fifty different immunoglobulin structures with $\beta$-sheet
architecture (left), the second set consisted of fifty hemoglobins with
$\alpha$-helical structures (middle) and, finally, the third set comprised fifty different myoglobin structures which also have $\alpha$-helical architecture (right).

While the computation of $Q_l$ and $\hat{W}_l$ in liquids is straight forward, for their correct calculation in proteins one needs to consider the following points: (1) The size of the protein must be large enough to obtain meaningful average values. Because helices or $\beta$-sheets do not have ``interior'' points, the notion of packing itself is meaningful only for tertiary structures. Hence, the proteins must contain enough tertiary ``structure'' \cite{Richards-94}; (2) A meaningful cut-off distance must be selected in the identification of near neighbors for a given side chain. The three protein data sets used in this study are sufficiently large that $Q_l$ and $\hat{W}_l$ can be easily computed to assess the degree of anisotropy in the relative residue-residue orientations. We used a neighbor ``cut-off'' distance that is 1.2 times the value of the position of the first peak in the radial distribution function for each structure. This definition ensured that all the side chains in the first coordination shell were counted as neighbors \cite{Steinhardt-83}. In the calculations presented here the centers of the residues were taken to be the geometric centers of the heavy atoms in the side chains.

\begin{figure}
\includegraphics{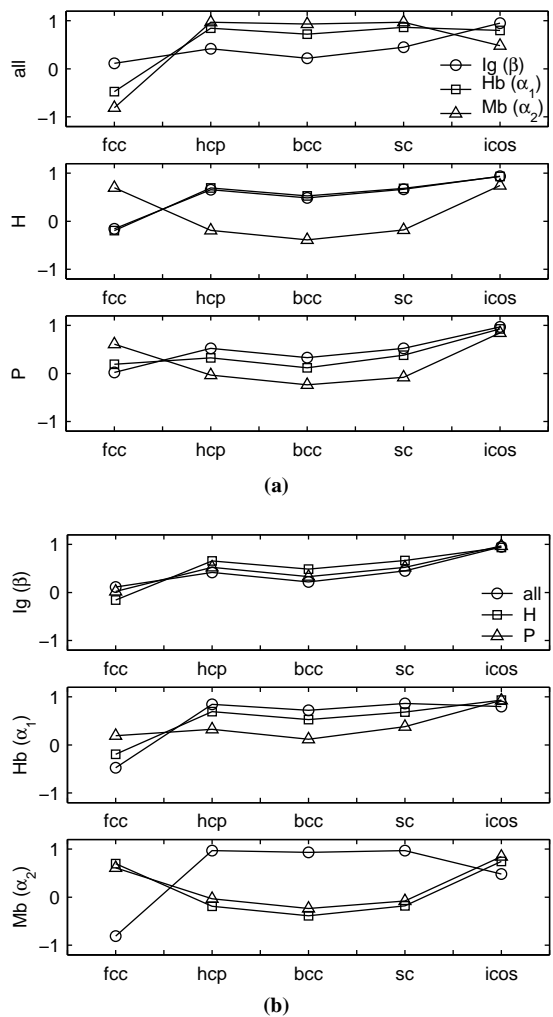}
\caption{\label{fig:QLMcorr} 
{{\bf (a)} Correlation coefficients between theoretical orientational order parameters (OOPs) calculated for simple cluster geometries and OOPs calculated for the three protein sets presented in Table~\ref{tab:table1}.
The results are shown by averaging: (i) over all the residue types (top, \emph{all}), (ii) over five most hydrophobic residues (middle, \emph{H}) and (iii) over the five most strongly polar side chain types (bottom, \emph{P}). The three curves in each figure correspond to correlation coefficients calculated for the first set of $\beta$-strand immunoglobulins (circles), for the second set of hemoglobins with prevalent $\alpha$-helical structures (squares), and for the third set of myoglobins with prevalent $\alpha$-helical structures (triangles). 
{\bf (b)} Same data as in Fig. \ref{fig:QLMcorr}a. Here, the results are shown for the values calculated for: (i) the set of immunoglobulins (top, $\beta$), (ii) the set of hemoglobins (middle, $\alpha_1$), and (iii) the set of myoglobins (bottom, $\alpha_2$). In each figure, the three curves correspond to correlation coefficients calculated by averaging over all the residue types  (circles), over the five most hydrophobic residues (squares), and over the five most strongly polar side chain types (triangles).}}
\end{figure}

Comparison of the results in Fig. \ref{fig:QLMs}a and Fig. \ref{fig:QLMs}b shows that a simple ``standard'' crystallographic type of order is absent in proteins. This shows that the interior of proteins does not have a simple point group symmetry. We calculated the values of the correlation coefficients between the OOPs presented in Fig. \ref{fig:QLMs}b and the ones in Fig. \ref{fig:QLMs}a.
These calculations show that a simple ``standard'' crystallographic type of order is absent in proteins. The interior of proteins does not have a simple point group symmetry, which would have permitted the development of a simple, quantitative description of side chains packing. In Fig. \ref{fig:QLMcorr} are shown the correlation coefficients between the theoretical OOPs values (Fig.~\ref{fig:QLMs}a) and the OOPs (Fig.~\ref{fig:QLMs}b) calculated for the three sets of proteins with $\alpha$-helical and $\beta$-strand structures (Table~\ref{tab:table1}) . 
The absolute magnitudes of the correlation coefficients are not very meaningful when comparing small data sets, but their \emph{relative} values can give us a quantitative indication if the order parameter values are closer to one type of cluster geometry (e.g. fcc) than to another (e.g. icos). 
The results are shown in Fig. \ref{fig:QLMcorr}a by averaging:(a) over all the residue types (top, \emph{all}), (b) over the five most hydrophobic residues (middle, \emph{H}) and (c) over the five most strongly polar side chain types (bottom, \emph{P}). Please note that the \emph{all} data represents averages computed over all 20 amino acids, while the \emph{H} and \emph{P} sets correspond to only 5 amino acids each. Therefore, the results for the \emph{all} set are not necessarily averages of the corresponding \emph{H} and \emph{P} values. The three curves in each figure correspond to correlation coefficients calculated for the immunoglobulins with $\beta$-sheet architecture (circles), for the set of hemoglobins with $\alpha$-helical structures (squares), and for the $\alpha$-helical myoglobins (triangles). A different plot of the same correlation coefficients is also shown in Fig. \ref{fig:QLMcorr}b for the values calculated for: (a) the set of immunoglobulins (top, $\beta$), (b) the set of hemoglobins (middle, $\alpha_1$), and (c) the set of myoglobins (bottom, $\alpha_2$). The three curves correspond for each figure to correlation coefficients calculated by averaging over all the residue types  (circles), over the five most hydrophobic residues (squares), and over the five most strongly polar side chain types (triangles).

	The analysis of the results in Fig. \ref{fig:QLMcorr} leads to the following conclusions: (1) There is no clear trend towards a single geometry associated to side chain packing. Nevertheless, we observe significant values for $Q_4$ and $Q_6$ that suggests that high order orientational correlations characteristic of fcc, bcc, or icosahedral (icos) clusters (or a combination of these) are observed in proteins with $\alpha$-helical and $\beta$-sheet architecture. (2) While there are strong correlations indicating an icosahedral packing of residues (most evident for polar residues in Fig. \ref{fig:QLMcorr}a, bottom), we also find significant contributions from the other types of cluster geometries. For example, for both $\beta$-strand (Fig. \ref{fig:QLMcorr}b top) and $\alpha$-helical (Fig. \ref{fig:QLMcorr}b middle) architectures, both the hcp and sc geometries make significant contributions. 
This is in accord with the observations of Bagci et al. \cite{Jernigan-02} who showed that there is a general uniform distribution of residues in proteins, two-thirds being approximately fcc packed and one-third occupying random positions. We also observed high correlations between the fcc packing and the orientational order of both hydrophobic and strongly polar residues in some $\alpha$-helical molecular structures like in myoglobins (Fig. \ref{fig:QLMcorr}a middle and bottom and Fig. \ref{fig:QLMcorr}b bottom), but there is also a high correlation for the icosahedral character of packing in those cases. (3) There is a strong similarity in the distribution of OOPs between the $\beta$-sheet immunoglobulin structures and the $\alpha$-helical hemoglobins (Fig. \ref{fig:QLMcorr}b). However, there is considerable difference in the orientational order between the two $\alpha$-helical proteins (hemoglobins and myoglobins). This suggests that even within a given fold there can be variations in the precise orientational registry of the side chains. Similarly, it appears that proteins with vastly different architecture can have similar orientational order. A more detailed investigation of orientational order in a large data set of proteins is warranted. 

The results presented here show that the bond-orientational order parameters are useful quantitative tools in investigating the orientational order of side chains in proteins. They also demonstrate that, for a given native state topology, the orientational packing of side chains may be decomposed into a linear combination of simple cluster geometries. 
The importance of orientational degrees of freedom in the packing of side chains reinforces the need for deciphering anisotropic inter-residue potentials from PDB structures. A method to extract such potentials is described next. 


\subsection{The Local Reference Frames for Amino Acids}
\label{sec:LRFs}

\begin{figure}
\includegraphics{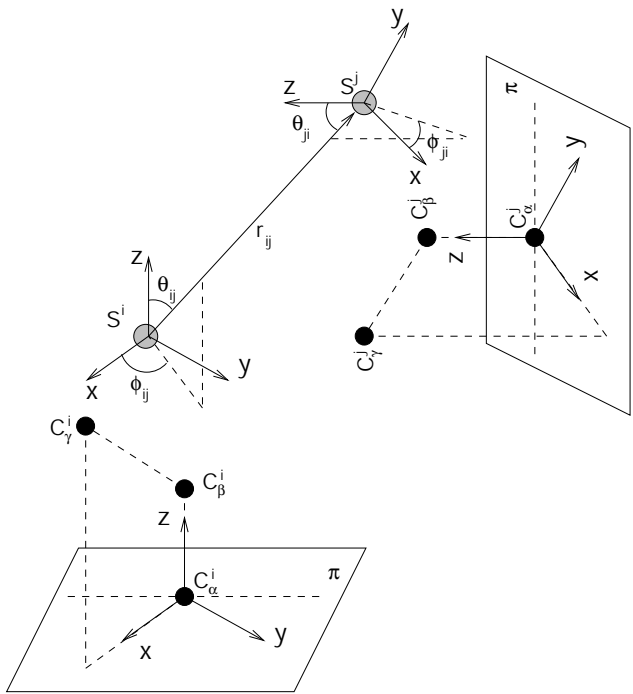}
\caption[]{\label{fig:LRFdef}
{The definition of a local reference frame (LRF) for residue-residue interactions. The local $Oz$ axis is defined by the positions of the $C_{\alpha}$ and $C_{\beta}$ atoms and the local $Ox$ axis is defined by the positions of the $C_{\beta}$ and $C_{\gamma}$ atoms. In all the cases, the local reference frames should be right handed. The interaction centers $S^{i}$ are defined as the centers of mass of the heavy atoms in the side-chains. The details are given in the text.}}
\end{figure}

\begin{figure}
\begin{center}
\includegraphics{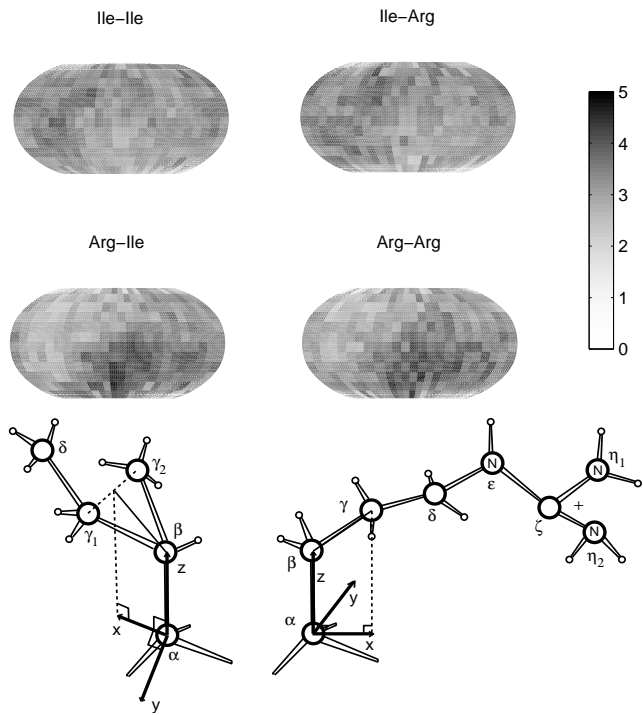}
\end{center}
\caption{\label{fig:IR2x2}
{Orientational probability density maps for Ile-Ile, Ile-Arg, Arg-Ile and Arg-Arg interactions. The probability amplitudes correspond to the scale shown in the scale bar, in units of $10^{-3}$. Extreme cases of highly hydrophobic (Ile) and the highly hydrophilic (Arg) residue are chosen to investigate if hydropathic properties are reflected in maps. Sample local reference frames (LRFs) for Ile and Arg are also depicted. The origins of the LRFs are placed in the centers of mass of the heavy atoms in the respective side-chains.}}
\end{figure}

As demonstrated in the previous sections, we need to define local reference frames (LRFs) for all the types of protein side chains in order to build a quantitative description of their relative orientations.
The definition of LRFs for SC-SC interactions is illustrated in Fig. \ref{fig:LRFdef}. 
The local $Oz$ axis is defined by the positions of the $C_{\alpha}$ and $C_{\beta}$ atoms. The local $Ox$ axis is defined by the positions of the $C_{\beta}$ and $C_{\gamma}$ atoms. The right handed nature of the LRFs determines automatically the direction of the $Oy$ axis if $Ox$ and $Oz$ are known. Once the local axes are defined, the polar angles $\theta$ and $\phi$ and the inter-residue distances $r$ can be used as internal degrees of freedom that describe the relative residue-residue orientations as shown in Fig. \ref{fig:LRFdef}.  
These definitions are altered for the following special cases: (1)~Gly does not have $C_{\beta}$ and the local $Oz$ axis is defined by the bisector of the angle defined by $N^{i}$, $C_{\alpha}^{i}$ and $C^{i}$, (2) both Gly and Ala do not have $C_{\gamma}$ and the local $Ox$ axis is defined as parallel to the direction defined by the backbone atoms $N^{i}$ and $C^{i}$, (3)~for Cys and Ser the corresponding coordinates of the S and O atoms are substituted for the coordinates of the missing $C_{\gamma}$, and (4)~the coordinates of the midpoint between the two $C_{\gamma}$ atoms are used to define the direction of the $Ox$ axis for Ile and Val.  For each amino acid the interaction center ($S^{i}$) is defined as the center of mass of the heavy atoms in the side chain with the exception of Gly for which the position of $S^{i}$ coincides with $C_{\alpha}$. Sample LRFs for Ile and Val are also shown in Fig. \ref{fig:IR2x2}.

\subsection{Finding the structure-derived potentials}
\label{sec:Pot}

From the definition of the LRFs, it follows that the SC-SC interaction potentials depend on five independent parameters that define the relative positions of two distant side-chains $i$ and $j$: $r_{ij}$, $\phi_{ij}$, $\theta_{ij}$, $\phi_{ji}$ and $\theta_{ji}$. We assume that these are the only geometrical factors that influence the two-body interactions between two residues $i$ and $j$ ($|j-i|\geq 4$). One more independent parameter (a torsional angle around $r_{ij}$) can be used for a complete description of the relative orientations of the two residues in three dimensions, but it is not employed in this work because its influence is expected to be important only for short range interactions. Accounting for this sixth angular parameter would also increase dramatically the statistical requirements for the protein database that is analyzed. Assuming pairwise interactions, the distance and orientation dependent potential for the residue pair $ij$ is
\begin{equation}
U^{ij} = U_{DO}^{ij}(r_{ij},\phi_{ij},\theta_{ij},\phi_{ji},\theta_{ji})
\end{equation}
We further assume that $U_{DO}^{ij}$ can be decomposed as
\begin{eqnarray}
U_{DO}^{ij}(r_{ij},\phi_{ij},\theta_{ij},\phi_{ji},\theta_{ji})&=&
U_{DO}^{ij}(r_{ij},\phi_{ij},\theta_{ij}) \nonumber\\ 
& &+ U_{DO}^{ji}(r_{ji},\phi_{ji},\theta_{ji})
\label{eq:Utot-1}
\end{eqnarray}

%
We use the $U_{DO}$ notation for the statistical potentials that are both \emph{distance} and \emph{orientation} dependent, and the $U_{D}$ notation for potentials that depend solely on inter-residue \emph{distances}. 
The assumption in Eq. \ref{eq:Utot-1} on the separability of potentials is not always valid. For a system of interacting side chain pairs, described by a Boltzmann equilibrium, Eq. \ref {eq:Utot-1} is consistent, however, with the probabilistic relation
\begin{eqnarray}
P_{total}^{ij}(r_{ij},\phi_{ij},\theta_{ij},\phi_{ji},\theta_{ji}) &=& 
P^{ij}(r_{ij},\phi_{ij},\theta_{ij}) \nonumber\\
& &\times P^{ji}(r_{ji},\phi_{ji},\theta_{ji})
\label{eq:probab}
\end{eqnarray}
where $P_{total}^{ij}(r_{ij},\phi_{ij},\theta_{ij},\phi_{ji},\theta_{ji})$ is the probability to find a pair of interacting side chains $i$ and $j$ separated by a distance $r_{ij}=r_{ji}$ between their interaction centers, and with relative orientations given by the set of ($\phi_{ij},\theta_{ij}$) angles in the $LRF_i$ frame, and by the set of ($\phi_{ji},\theta_{ji}$) angles in the $LRF_j$ frame (see Fig. \ref{fig:LRFdef}). Eq. \ref{eq:Utot-1} implies, therefore, that the relative orientations of the local reference frames $LRF_i$ and $LRF_j$ of two interacting side chains $i$ and $j$ do not depend on each other. As suggested by previous studies \cite{Sippl95,Dill-96,Jernigan-96}, this type of independence could be expected for side chains that are separated by a large enough number of peptide bonds along the backbone. This assumption is reasonable in our analysis because only side chains that are separated by at least five peptide bonds along the protein backbone are considered. This corresponds to residues that are found on a ``topological level'' $k\geq5$ \cite{Sippl90}. Besides this constraint, for simplicity, we are also considering (as in \cite{Dill-96}) that the SC-SC interactions are independent on sequence separation (i.e. for $k\geq5$, all side chains are on the same topological level).

\subsubsection{The Boltzmann Device}

The Boltzmann device assumes that the known protein structures from protein databases (such as PDB) correspond to classical equilibrium states. The SC-SC potentials can therefore be related to position probability densities $f(r)$ (see Eq.~\ref{eq:boltzmann}), where $r$ can be the radial distance or the angle between the side-chains \cite{Sippl95}. In many studies, $f(r)$ can be replaced by the normalized pair distribution functions $g(r)$ such that
\begin{equation}
U_D^{ij}(r)=-kT\ ln\left[\frac{g^{ij}(r)}{g_{ref}(r)}\right]
\end{equation}
for the distributions depending only on distances.
We adopt a more general treatment that defines
\begin{equation}
U_{DO}^{ij}(r,\phi,\theta)=-kT\ ln\left[\frac{P^{ij}(r,\phi,\theta)}{P_{ref}(r,\phi,\theta)}\right]
\end{equation}
for the distance and orientation-dependent distributions. To be consistent with previous studies, we consider the reference pair distribution functions $P_{ref}$ to be the corresponding radial or angular pair distributions that are obtained through an analysis of all twenty residue types. A database of non homologous proteins can be used to estimate the pair distributions and to extract amino acid specific interaction potentials that are consistent with a large set of protein structures. 

	An important issue that appears when using probability density functions with the Boltzmann device for constructing statistical potentials is ``the problem of small data sets''. As noted by Sippl \cite{Sippl90}, dividing the SC-SC pair frequencies by both side chain type and distance intervals results in situations when the available data is too small for conventional statistical procedures. This problem was solved by Sippl by proposing a ``sparse data correction'' formula that builds the correct probability densities as linear combinations between the measured data and the reference, total probability densities obtained by averaging over all twenty SC types. For the general, orientation-dependent probability densities the sparse data correction can be written as
\begin{eqnarray}
P^{ij}_{corr}(r,\phi,\theta) &=& 
\frac{1}{1+m'\sigma} P_{ref}(r,\phi,\theta) \nonumber \\
& &+ \frac{m'\sigma}{1+m'\sigma} P^{ij}(r,\phi,\theta)
\label{eq:sdc}
\end{eqnarray}
where $P^{ij}$ are the actual probability densities obtained from the database for the $ij$ pair of side chains,  $P^{ij}_{corr}$ are the corrected probabilities, and $P_{ref}$ is the reference probability density. A modification introduced by the orientational dependence in our case is that the number of measurements $m$ becomes $m'=m/sin(\theta_k)$, as $k$ equiangular intervals are used for the $\theta$ angle. This is necessary for accounting for the azimuthal dependence of volume elements in spherical coordinates.
The $\sigma$ parameter is a constant that controls how many actual measurements $m'$ must be observed so that both the actual probabilities and the reference would have equal weights. As in other studies, we used the value $\sigma=1/50$ \cite{Sippl90,Hendlich,Dill-96}.

\subsubsection{Residue-Residue Interaction Probability Maps}

\begin{figure}
\includegraphics{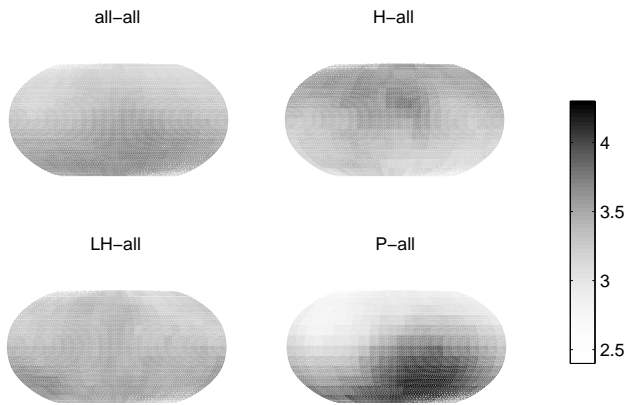}
\caption[]{\label{fig:HP2x2}
{Orientational probability density maps for all-all, H-all, LH-all and P-all interactions. Here ``all'', ``H'', ``LH'', and ``P'' are virtual residues. The ``all'' type is obtained by averaging over all the observed SC-SC orientations. ``H'' is obtained by averaging over the five highly hydrophobic amino acids (Ile, Val, Leu, Phe, Met), ``LH'' corresponds to the average of eight less hydrophobic residues (Ala, Gly, Cys, Trp, Tyr, Pro, Thr, Ser), and ``P'' is obtained by averaging over seven highly hydrophilic amino acids (His, Glu, Asn, Gln, Asp, Lys and Arg)~\cite{Biochem-1,Eisenberg-84}. The hydropathic properties and effects due to the finite size of proteins are clearly reflected in these maps. The probability amplitudes correspond to the bar scale shown, with units of $10^{-3}$.}}
\end{figure}

Using our proposed definitions of residue-specific local reference frames (introduced in section \ref{sec:LRFs}) and the histogram extraction procedure (described in detail in Appendix \ref{Appx:HistEx}), we constructed radial and orientational SC-SC interaction probability maps. All the combinations of side chain types were investigated. From the normalized angular histograms, the resulting probability density maps for residue-residue interactions were calculated. 
For purpose of illustration, the data shown in all the probability density maps presented in this paper were obtained by averaging over the entire distance range for which the radial and angular histograms were constructed (2\AA\ to 26\AA), as explained above.

The orientational probability maps for interactions between Ile (one of the most hydrophobic residues) and Arg (highly hydrophilic) are shown in Fig.~\ref{fig:IR2x2} together with a schematic representation of their associated LRFs. The LRFs in Fig.~\ref{fig:IR2x2} are shown in the proximity of the $C_{\alpha}$, to illustrate the computational process of constructing them for actual structures. However, for calculations the LRFs are translated such that their origin corresponds to the geometrical center of the heavy atoms in the side chains (i.e. the interaction centers $S_i$). The orientational anisotropy is stronger around Arg than around Ile (Fig.~\ref{fig:IR2x2}), which is possibly due to the relatively longer Arg side chain. Lys, another strong hydrophilic residue, presents a relative orientational interaction probability map similar to Arg but with an even stronger anisotropic character. The similarity between the Ile-Ile and Ile-Arg maps and between the Arg-Ile and Arg-Arg is largely a reflection of the fact that the statistical data collected here does not depend on the relative orientation of the second side chain ($SC_{j}$) in the reference system of the first SC ($LRF_{i}$). This finding may be a consequence of the assumptions behind Eq.~\ref{eq:probab}.

The most noticeable difference between the maps calculated in the LRFs of Ile and Arg is that the relative amplitudes of the interaction probabilities at the ``poles'' are reversed. In other words, the most preferred interaction loci for Ile is around its ``north pole'' (i.e. towards the positive $Oz$ axis, away from the local backbone) while for Arg is in the proximity of its ``south pole''. To investigate if this feature is due to the hydrophobic/hydrophilic character, we computed average orientational interaction probability maps for three groups of residues: (a) five highly hydrophobic (H) residues Ile, Val, Leu, Phe, Met, (b) eight less hydrophobic (LH) residues Ala, Gly, Cys, Trp, Tyr, Pro, Thr, Ser, and (c) seven highly hydrophilic ($P$) residues His, Glu, Asn, Gln, Asp, Lys and Arg. This classification was suggested by various hydropathy scales \cite{Biochem-1,Eisenberg-84}. The computed average maps are presented in Fig.~\ref{fig:HP2x2} together with the maps for the average virtual residue ``all''.    

The ``all-all'' map shows that, for any residue there is a slightly higher probability to find more residues along the attached negative $Oz$ axis in their respective LRFs than in the ``northern hemisphere''. The definition of the LRFs was made such that the positive $Oz$ direction points away from the nearest backbone atoms. Due to the finite size and the relatively compact three-dimensional shapes of most proteins analyzed here, we do expect to find less residues in that direction for higher SC-SC distances and, therefore, at larger distances from the backbone. However, the ``H-all'' map (Fig.~\ref{fig:HP2x2}) shows that the highly hydrophobic amino acids present the reverse situation: a higher average probability to find other residues at the ``north pole'' of their side chains. This arises because such residues are mostly found in the ``interior'' of proteins, ``protected'' from water. The same arguments apply to the observation that, on average, the highly hydrophilic residues have preferential coordination loci shown on the ``P-all'' map in the proximity of their ``south pole'', as it is expected that their positive $Oz$ axis points more often toward the water molecules. It is reassuring that the average over the less hydrophobic residues, shown in the ``LH-all'' map of Fig.~\ref{fig:HP2x2}, does not reflect the existence of significant orientational preferences.

\section{Results: How important is the information on relative residue-residue orientations?}

\begin{figure}
\includegraphics{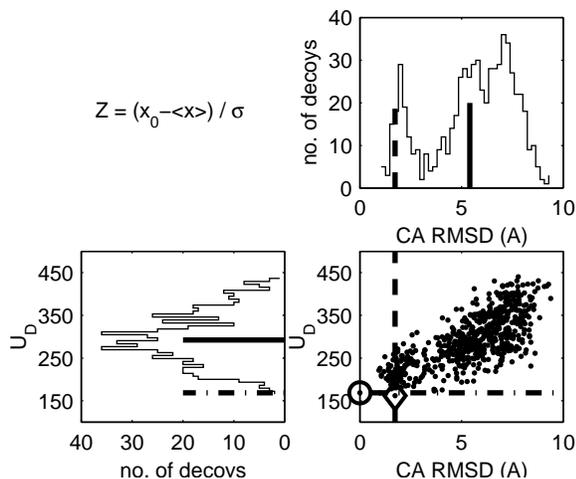}
\caption{\label{fig:Zdef-1-S-EX}
{The $RMSD$ and energy Z-scores for decoy sets of protein structures. The distance-dependent energy ($U_D$) for the set of 654 decoys of Calbindin (3icb), is plotted as a function of the $C_{\alpha}$ RMSD. The Z-scores are proportional to the distance of the corresponding parameter of interest (depicted as interrupted lines) from the mean values of their distributions (solid lines), in units of standard deviations $\sigma$.}}
\end{figure}

	We performed tests using the distance and orientation dependent statistical potentials ($U_{DO}$) as scoring functions and we assessed their performance on a standard database of decoys developed by Samudrala and Levitt \cite{Samudrala}. The results are compared with respect to the performance of potentials dependent solely on distance ($U_D$). Similar decoy tests have been shown to be useful in analyzing the ability of various potential schemes to correctly recognize the native state \cite{Vajda-00,Kollman-01a}. There are two main aspects that need to be taken into consideration when using decoy protein structures for tests: (1)~The energy of the native state should be as low as possible, and (2)~The decoy structure that has the lowest root mean square $C_{\alpha}$ distance (RMSD) from the native state should also have a low energy \cite{Vajda-00,Kollman-01a,Kollman-01b}. 
For studying these aspects from a quantitative point of view, we use the Z-score for both the energy ($Z_{E}$) and the RMSD deviations of the decoy structures ($Z_{RMSD}$). The definition of these two types of Z-scores is illustrated in Fig. \ref{fig:Zdef-1-S-EX} where the distance-dependent energies ($U_D$) that we computed for the set of 654 decoys of Calbindin (3icb), is plotted as a function of the $C_{\alpha}$ RMSD, calculated with respect to the native state. 
The Z-score of a statistical quantity $x$ is 
\begin{equation}
Z=\frac{x-\overline{x}}{\sigma}
\end{equation}
where $\sigma$ is the standard deviation and $\overline{x}$ is the mean of the distribution of $x$ values. In our case, we use $x$ both as the energy and the RMSD of the set of decoys. In all cases, a good scoring function will lead to a negative Z-score. In Fig.~\ref{fig:Zdef-1-S-EX}, we show the radial energy term computed using our method for the Calbidin decoy set (3icb in \cite{Samudrala}). The high similarity between this plot and the energies depicted in Fig. 1 of Samudrala and Levitt \cite{Samudrala} confirms that the present radial potentials are essentially the same as those used by other groups. In this figure, the circle shows the position of the native state and the diamond shows the position of the decoy structure that has the lowest energy. For a potential (or scoring function) that is efficient in discriminating the native state of a protein from a set of decoys it is expected that (1) the native state (circle) corresponds to the lowest interaction energy, and (2) the decoy with the lowest energy (diamond) should have as small an RMSD as possible. Both criteria are important and we find that both the $Z_E$ and $Z_{RMSD}$ scores defined here are useful, quantitative measures of the performance of the potential energy function. These Z-scores are proportional to the distance of the corresponding parameter of interest (depicted in Fig.~\ref{fig:Zdef-1-S-EX} as interrupted lines) from the mean values of their distributions (solid lines), in units of standard deviations $\sigma$.

	Due to the large number and diversity of decoy sets that are employed in this paper, we do not repeat the specific description of the methods used for generating each decoy set and of their names (e.g. ``single'',``lmds'', ``fisa'', etc.). That information is provided by the ``Decoys `R' Us'' database (http://dd.stanford.edu) and by the corresponding publications \cite{Samudrala}. Details are provided for the decoys that are relevant to the results obtained for our tests.

\begin{figure*}
\includegraphics{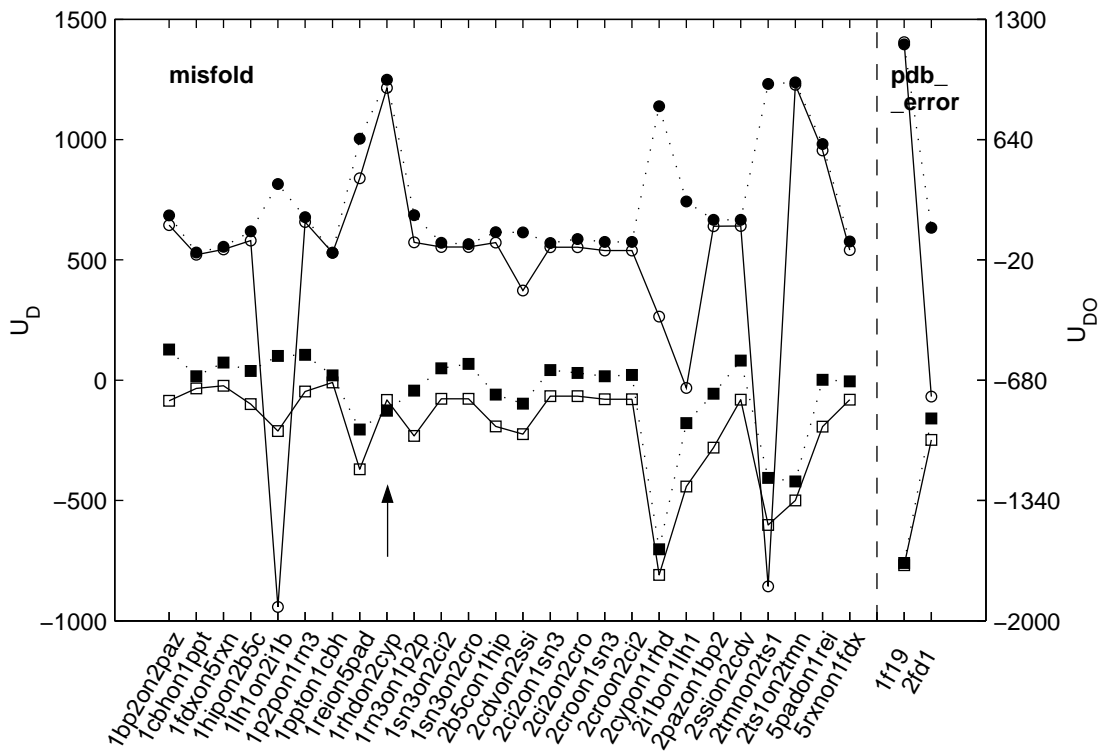}
\caption{\label{fig:fsingle}
{Results of tests on the ``single'' decoy sets \cite{Samudrala} that consist of pairs of correct (i.e. native) and incorrect structures. The plot compares the total statistical potential values obtained for distance-only dependent potentials ($U_{D}$, squares), and for distance and orientation dependent potentials ($U_{DO}$, circles). 
The points united by dotted lines (filled symbols) correspond to incorrect structures, while the continuous lines join scores obtained for correct states (open symbols). The arrow emphasizes a case where the $U_{D}$ score fails to identify the native state, while $U_{DO}$ succeeds. These types of tests, concerned with discriminating the native state from a single decoy, are relatively easy and both the $U_{D}$ and $U_{DO}$ scores succeed to identify the ``correct'' structures in a majority of cases.}}
\end{figure*}

	In Fig. \ref{fig:fsingle} are compared total statistical potential values obtained for distance-only dependent potentials ($U_{D}$, squares), and for distance and orientation dependent potentials ($U_{DO}$, circles) for decoy structures in the ``single'' sets (``misfold'' and ``pdb\_error'') of the ``Decoys `R' Us'' database \cite{Samudrala}. 
\begin{figure*}
\includegraphics{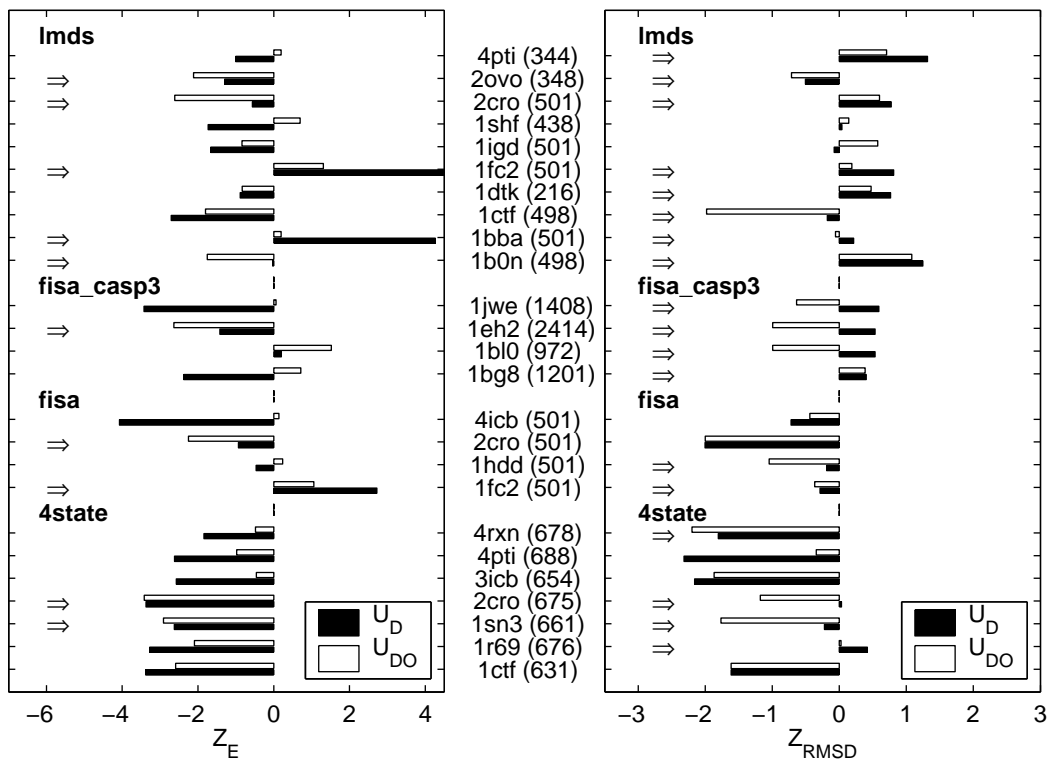}
\caption{\label{fig:ZEZR-4state}
{The energy ($Z_E$) and $C_{\alpha}$ RMSD Z-scores ($Z_{RMSD}$) calculated for the multiple decoy sets ``lmds'', ``fisa\_casp3'', ``fisa'' and ``4state'' \cite{Samudrala}. The numbers in brackets represent the number of decoys in each set, including the native structure. Z-scores calculated using statistical potentials dependent only on distance $U_{D}$ (dark bars) and distance \emph{and} orientation dependent potentials ($U_{DO}$, white) are compared. More negative Z-scores are better, and the cases in which $U_{DO}$ gives better results than using $U_{D}$ are emphasized by the arrows on the left. When both $Z_E$ and $Z_{RMSD}$ Z-scores are considered, the inclusion of orientational information improves the performance in a majority of cases.}}
\end{figure*}
\begin{figure*}
\includegraphics{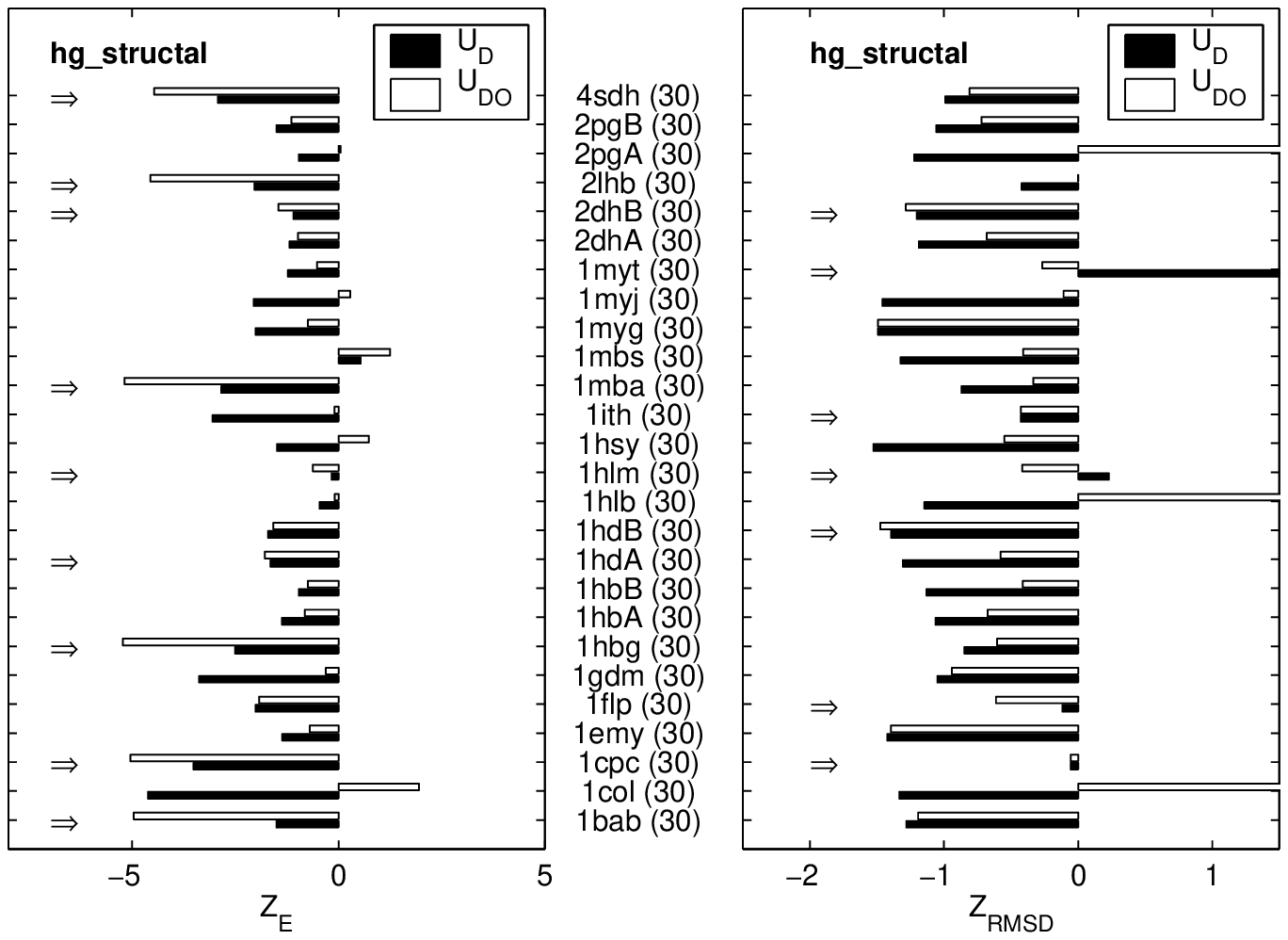}
\caption{\label{fig:ZEZR-hg}
{Energy ($Z_E$) and $C_{\alpha}$ RMSD Z-scores ($Z_{RMSD}$) calculated for the multiple decoy sets ``hg\_structal'' \cite{Samudrala}. The numbers in brackets represent the number of decoys in each set, including the native structure. Z-scores calculated using statistical potentials dependent only on distance ($U_{D}$, dark bars) and distance \emph{and} orientation dependent potentials ($U_{DO}$, white bars) are compared. More negative Z-scores are better. The cases in which $U_{DO}$ gives better results than using $U_{D}$ are emphasized by the arrows on the left.}}
\end{figure*}
\begin{figure*}
\includegraphics{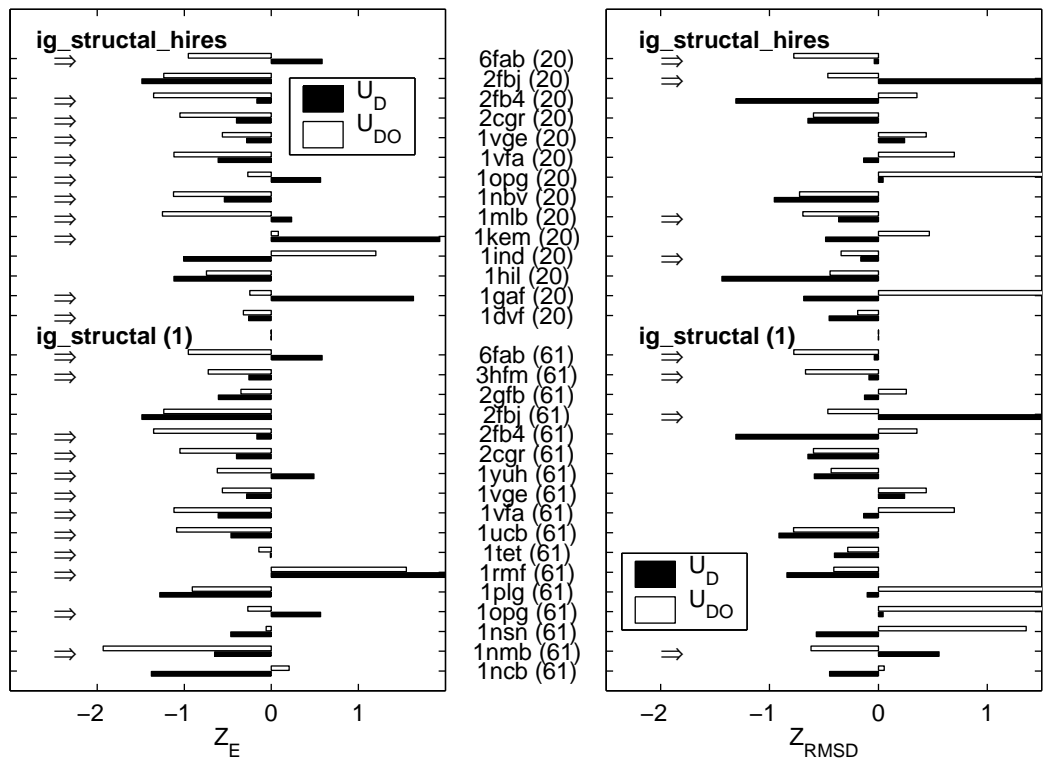}
\caption{\label{fig:ZEZR-igA}
{Energy ($Z_E$) and $C_{\alpha}$ RMSD Z-scores ($Z_{RMSD}$) calculated for the multiple decoy sets ``ig\_structal'' ($1^{st}$ part) and ``ig\_structal\_hires'' \cite{Samudrala}. The numbers in brackets represent the number of decoys in each set, including the native structure. Z-scores calculated using statistical potentials dependent only on distance ($U_{D}$, dark bars) and distance \emph{and} orientation dependent potentials ($U_{DO}$, white bars) are compared. More negative Z-scores are better. The cases in which $U_{DO}$ gives better results than using $U_{D}$ are emphasized by the arrows on the left.}}
\end{figure*}
\begin{figure*}
\includegraphics{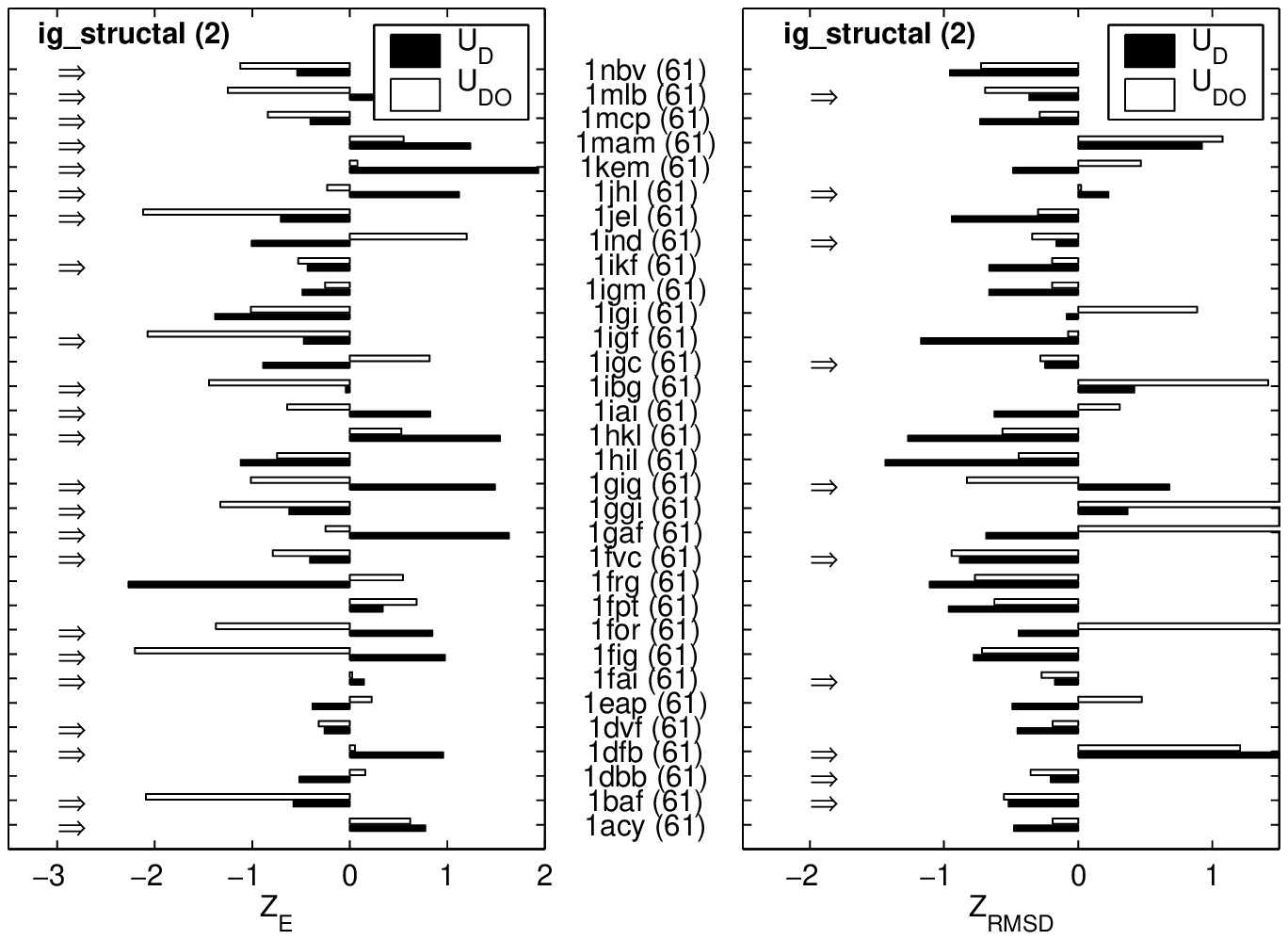}
\caption{\label{fig:ZEZR-igB} 
{Energy ($Z_E$) and $C_{\alpha}$ RMSD Z-scores ($Z_{RMSD}$) calculated for the multiple decoy sets ``ig\_structal'' ($2^{nd}$ part) \cite{Samudrala}. The numbers in brackets represent the number of decoys in each set, including the native structure. Z-scores calculated using statistical potentials dependent only on distance ($U_{D}$, dark bars) and distance \emph{and} orientation dependent potentials ($U_{DO}$, white bars) are compared. More negative Z-scores are better. The cases in which $U_{DO}$ gives better results than using $U_{D}$ are emphasized by the arrows on the left. It is observed that in a majority of cases the inclusion of orientational information improves the performance of both $Z_E$ and $Z_{RMSD}$ scores.}}
\end{figure*}

The ``misfold'' set contains, for each native protein, an alternative structure generated by placing the same sequences on different folds with the same number of residues (e.g. the ``1bp2on2paz'' notation means that the ``1bp2'' sequence has been placed on the ``2paz'' fold \cite{Samudrala}). The incorrect structures in the ``pdb\_error'' set are experimental structures that have been substantially re-refined or found to contain errors. 
The points united by dotted lines (filled symbols) correspond to incorrect structures, while the continuous lines (open symbols) join scores obtained for correct states (i.e. ``native'' for the ``misfold'' set, and ``latest'' for the ``pdb\_error'' set). The arrow in Fig. \ref{fig:fsingle} emphasizes the case where the $U_{D}$ score fails to identify the native state, while $U_{DO}$ succeeds. Overall, the ``single'' decoy sets proved to be relative easy tests for both the $U_{D}$ and $U_{DO}$ scores. The orientation dependent $U_{DO}$ scores consistently result in better (i.e. smaller) values for the correct protein structures in all cases studied.

In all the tests performed, some of the decoy sets that were not appropriate for analysis were eliminated. For example, non-standard side chains were present in the decoys or in the native structures, for which no $U_{DO}$ potentials were constructed. Another reason for eliminating a few structures was the absence of enough heavy atom coordinates in some of their large side chains. This prevented the construction of LRFs for those side chain and, therefore, the correct estimation of their $U_{DO}$ potentials.

Much more challenging decoy tests than the ones presented above for the ``single'' decoys can be performed using the ``multiple'' decoy sets from the ``Decoys `R' Us'' database \cite{Samudrala}. The main goal of these tests is to identify the native structure from a set of many (i.e. tens, hundreds, or even thousands) of decoys. We discuss next the results of these tests using the $Z_E$ and $C_{\alpha}$ RMSD Z-scores defined as shown in Fig. \ref{fig:Zdef-1-S-EX}. From the ``multiple'' decoy sets available in the ``Decoys `R' Us'' database, the ``lattice'' set was eliminated from the analysis presented here because the new $U_{DO}$ statistical potentials used in this paper were constructed using a training set of real, non-homologous protein structures. The specific constraints imposed by the tetrahedral lattice topology on side chain conformations for the ``lattice'' decoys affect more negatively the performance of the $U_{DO}$ potentials than the performance of the distance-only dependent scores, $U_D$. 

	In Fig. \ref{fig:ZEZR-4state} are shown energy ($Z_E$) and $C_{\alpha}$ RMSD Z-scores ($Z_{RMSD}$) calculated for the multiple decoy sets ``lmds'', ``fisa\_casp3'', ``fisa'' and ``4state'' \cite{Samudrala}.
Z-scores calculated for the distance-only dependent statistical potentials $U_{D}$ (dark bars) and for the new $U_{DO}$ potential (white bars) values are compared. In Fig. \ref{fig:ZEZR-4state}, and in all the subsequent figures presenting Z-scores, more negative values are better, and the cases in which $U_{DO}$ gives better results than using $U_{D}$ are emphasized using arrows. It is observed that in a large number of cases the inclusion of orientational information improves the performance of both $Z_E$ and $Z_{RMSD}$ scores. An especially interesting case is the one of the ``lmds'' set in which all-atom distance dependent scores were shown to perform poorly \cite{Samudrala, Samudrala-Moult}. In this case, when both the $Z_E$ and $Z_{RMSD}$ Z-scores are considered, the new distance- and orientation-dependent potentials $U_{DO}$ performed much better than the distance-only dependent $U_D$ in a majority of cases.

In Figs. \ref{fig:ZEZR-hg}-\ref{fig:ZEZR-igB} are shown the corresponding results for the ``hg\_structal'', ``ig\_structal\_hires'' and ``ig\_structal'' decoy sets of the ``Decoys `R' Us'' database.
For the ``hg\_structal'' decoy sets of globins, improvements are observed in about a third of the test cases. 
However, for both the ``ig\_structal\_hires'' and ``ig\_structal'' decoy sets of immunoglobulins, more than 70\% of tests present an improvement in the performance the energy Z-scores when $U_{DO}$ potentials are employed ($Z_{E}(U_{DO})$).

	The results for all the sets of ``multiple'' decoy tests are summarized in Table \ref{tab:table2}. 
As mentioned above, particularly strong improvements are observed when using $U_{DO}$ for the ``lmds'' set, for which distance-dependent scoring functions were reported before to have a weak performance \cite{Samudrala,Samudrala-Moult}. The very good energy Z-scores ($Z_E(U_{DO})<Z_{E}(U_{D})$ for more than $70\%$ of the decoy sets) that were obtained for the large decoy sets of immunoglobulins (``ig\_structal'' and ``ig\_structal\_hires'') suggest that considering the orientational dependence ($U_{DO}$) in the cases of proteins with a significant $\beta$-sheet architecture it is more important than for proteins that are mainly $\alpha$-helical (e.g. the ``hg\_structal'' sets).
At the same time, the fact that $U_{DO}$ improves both Z-scores in about a third of the ``hg\_structal'' decoy tests, could be explained by the globular shape of these predominantly $\alpha$-helical structures. These results agree to the rather intuitive observation that details about the relative side chain orientations could play a more significant role in compact, globular proteins than in proteins with relatively extended chains. Overall, it is shown that, for a majority of test cases, the $Z_E$ and the $Z_{RMSD}$ Z-scores are improved by including the relative SC-SC orientations in the structure analysis. This demonstrates that the performance of the statistical potentials in discriminating the native state is significantly enhanced by the inclusion of orientation dependent information.

	The results of the Z-score calculations for the decoy sets analyzed in this paper show that there are cases where the more detailed, distance- and orientation-dependent potentials ($U_{DO}$) do not provide a better performance in discriminating the native state as compared to using simple distance-dependent potentials ($U_{DO}$). Since both the $U_{DO}$ and $U_{D}$ potentials employed here were extracted from the same protein database and for the same number of radial bins, it is expected that the $U_{DO}$ values, while more detailed, are at the same time more prone to statistical errors than the $U_{D}$ potentials. The sparse data correction procedure ensures that reasonable values are employed for situations when little or no specific statistical data on relative residue-residue distances and orientations is available. However, this procedure alone can only ensure that realistic, accurate statistical potentials are obtained when a very large training database of non-homologue protein structures is employed. The relatively reduced size of the database of structures  employed here (and in \cite{Unres-1}) for extracting the statistical data could be an important factor responsible for the reduced $U_{DO}$ performance in some cases. 
\begin{table}
\caption{\label{tab:table2}
{Results of tests for the ``multiple'' decoy sets \cite{Samudrala}.}}
\begin{ruledtabular}
\begin{tabular}{ccc}
The ``multiple'' & $Z_{E}(U_{DO})$ & $Z_{RMSD}(U_{DO})$ \\
decoy sets\footnote{The $U_{DO}$ potentials were not ``trained'' for the lattice topology and, therefore, the ``lattice'' set \cite{Samudrala} was excluded from these tests (see text).} & $<Z_{E}(U_{D})$ & $<Z_{RMSD}(U_{D})$ \\
\hline
lmds & 50.0\% & 80.0\% \\       
fisa\_casp3 & 25.0\% & 100.0\% \\
fisa & 50.0\% & 50.0\% \\
4state & 28.6\% & 57.1\% \\
hg\_structal & 34.6\% & 26.9\% \\
ig\_structal\_hires & 78.6\% & 28.6\% \\ 
ig\_structal & 71.4\% & 28.6\% \\
\end{tabular}
\end{ruledtabular}
\end{table}

\begin{figure*}
\includegraphics{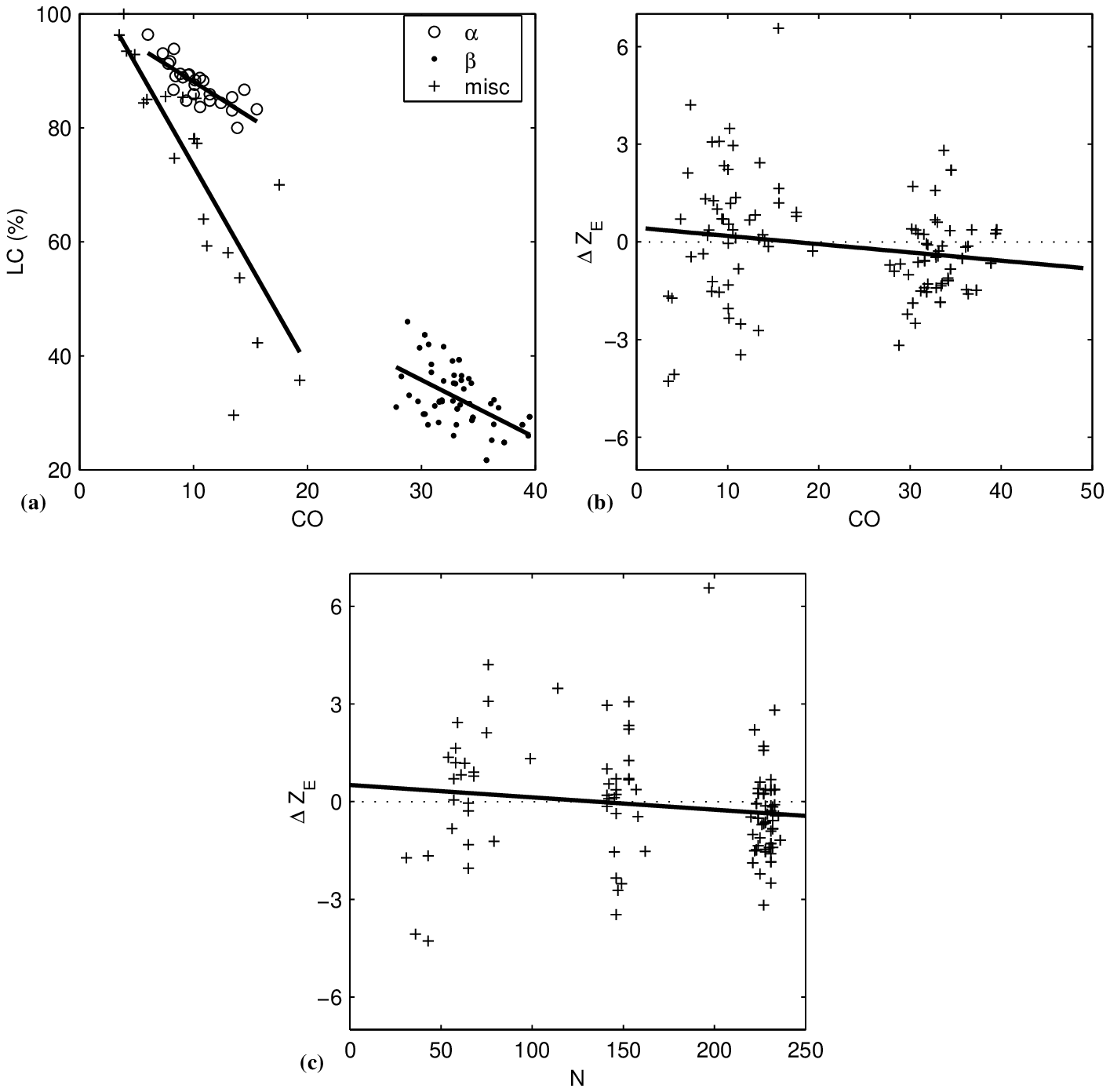}
\caption{\label{fig:ZER-fit}
{{\bf (a)} The dependence of the percent of local contacts (LC) on the contact order (CO) of the native structures for all the decoy sets analyzed in this paper. {\bf (b)} The difference $\Delta Z_{E}=Z_{E}(U_{DO})-Z_{E}(U_{D})$ for the energy Z-scores plotted versus the CO of the native state for each decoy set. {\bf (c)} $\Delta Z_{E}$ plotted versus the number of residues (N) for each decoy set. Negative $\Delta Z_{E}$ values correspond to better performing scores for the distance- and orientation-dependent potentials ($U_{DO}$). The lines, representing linear fits of $\Delta Z_E$, are shown for emphasizing the trends.}}
\end{figure*}

	The type of protein architecture and the number of residues are also factors that could influence the relative performance of the $U_{DO}$ and $U_{D}$ potentials. A useful quantitative parameter that is directly related to the type of protein architecture is the contact order (CO) \cite{Dill-93,Baker-98,Dima-02} of a protein structure. The CO definition employed here is
\begin{equation}
CO=\frac{1}{N_c}\sum^{N_c}_{<i,j>}\Delta S_{i,j}
\label{eq:CO}
\end{equation}
where $N_c$ is the total number of contacts and $\Delta S_{i,j}$ is the sequence separation, in residues, between contacting residues $i$ and $j$. Two residues are considered to be in contact when $|i-j|>1$ and any of the heavy atoms of residue $i$ is within 3.75 \AA\ of any of the heavy atoms of residue $j$ \cite{McCammon-02}. When the CO is small the protein structure presents mostly local contacts (i.e. $|i-j|<6$), and when CO is larger, the contacts are nonlocal \cite{Dill-93,Dima-02}. As shown in Fig. \ref{fig:ZER-fit}a we observed an inverse proportionality between the CO values calculated for all the native states of the decoys analyzed in this paper and the fraction of local contacts for various protein architectures. 
Based on this observation, we investigated the relative performance of the $U_{DO}$ and $U_{D}$ potentials as a function of CO values (Fig. \ref{fig:ZER-fit}b) and sequence length (Fig. \ref{fig:ZER-fit}c).
In Figs. \ref{fig:ZER-fit}b and \ref{fig:ZER-fit}c is shown the dependence of the difference $\Delta Z_{E}=Z_{E}(U_{DO})-Z_{E}(U_{D})$ for the energy Z-scores on contact order values (CO) and sequence length of the native state (N). Negative $\Delta Z_E$ values correspond to better performing scores for the distance- and orientation-dependent potentials ($U_{DO}$). The lines represent linear fits. While the trends are not very strong, it is observed that, for the energy Z-scores the novel $U_{DO}$ statistical potentials perform better then the $U_D$ potentials for longer proteins, presenting large contact orders, such as the $\beta$-sheet structures from the ``ig\_structal'' and ``ig\_structal\_hires'' sets.
It is observed that the inclusion of the information on relative SC-SC orientations makes the current version of the $U_{DO}$ statistical potentials more useful than $U_D$ for relatively large, globular proteins ($N>100$ residues), with a significant content of $\beta$-sheet structure and nonlocal contacts. 
At the same time, it is also observed that even for relatively short ($N<100$ residues) and $\alpha$-helical proteins (low CO values), both the $Z_E$ and $Z_{RMSD}$-scores can be improved when using $U_{DO}$ potentials but for a smaller number of cases. These results suggest that details about side chain - backbone interactions should be included in statistical potentials for short or $\alpha$-helical proteins with a high content of local contacts.

\section{Conclusions}

	Our results demonstrate that it is possible to improve the performance of energy based scoring functions by using statistical information extracted from the relative residue-residue orientations. Calculations of orientational order parameters for investigating the residue packing in proteins showed that architecture dependent packing is best described by a linear combination of simple cluster geometries (fcc, icos, hcp and bcc). This result reinforces the need for extracting orientation dependent potentials using PDB structures. We have defined a novel local reference system for each residue for quantitatively describing the relative three-dimensional orientations in a manner that is independent of the neighboring amino acids. Arguments based on experimental resolution of protein structures were used to derive the optimal bin size employed for collecting statistical data with both angular and distance dependence. The orientational information has both common and complementary significance as compared to the information that can be extracted from the relative residue-residue distances alone. The tests that we performed on a standard data base of artificially generated decoy structures suggest that this complementarity can be very important for a large class of protein structures. 
The novel distance and orientation dependent statistical potentials were shown to present an enhanced ability to recognize native-like protein folds, especially for larger structures with high contact orders (e.g. immunoglobulins). They should find use in constructing the next generation of coarse grained off-lattice protein simulations. These new potentials could also be instrumental in developing more efficient computational methods for structure prediction on much larger scales than it is currently possible, addressing one of the major goals of proteomics. To achieve this it will be important to also include the relative orientations between side chains and the protein backbone \cite{Keskin98F&D}.

\section{Acknowledgments}

	We thank Ruxandra I. Dima, Alan van Giessen and Tom Keyes for useful discussions. This work was supported by the National Institutes of Health R01 NS41356-01 (JES and DT), the National Science Foundation CHE-9975494 (JES), and CHE-0209340 (DT). The authors are grateful to the Center for Scientific Computing and Visualization at Boston University for computational resources. The data visualization was carried out using VMD \cite{VMD} and Matlab (The Mathworks, Inc., Natick, MA).

\appendix*

\section{Histogram Extraction}
\label{Appx:HistEx}

Previous studies suggest that only protein structures that are determined with a resolution of 2\AA\ or better should be used in the computation of $g^{ij}(r)$ \cite{Unres-1} from protein databases \cite{PDB}. 
A resolution of 2\AA\, for protein structures that are determined by X-ray crystallography corresponds to an accuracy of +/- 0.2\AA\ in atomic positions \cite{Moody}. The 2\AA\  resolution is often good enough to accurately assign hydrogen bonds and to allow for a limited interpretation of solvent structure.

The high resolution structural data are binned for the construction of either radial or orientation dependent pair distributions $g^{ij}(r)$ and $g^{ij}(\phi,\theta)$. It is important, therefore, to identify the minimal radial and angular bin sizes that ensure a certain high level of confidence that the information has been analyzed correctly and that no important artifacts due to the limitations of the experimental methods have been overlooked. In building a statistical distribution, the data that is collected from protein structures \cite{PDB} consists of inter-atomic distances. We expect that such data has an accuracy of $dR= +/- 0.4$\AA. 

For the radial dependence of the data, let the bin size be $L$. In the process of assigning a certain pair distance to a certain bin, if that value is exactly at the bin boundary, we have a probability of correct assignment $P_{ca}=0.5$. On the other hand, if the value is exactly in the middle of the bin, this probability is maximum (if $L\geq 2dR$ than $P_{ca}=1$). We estimate that the probability of correct radial assignment depends on the bin size $L$ and the accuracy $dR$ as 
\begin{equation}
P(L,dR)=2/L\ \int_{0}^{L/2} P_{ca}(x,dR) dx
\label{eq:Pca}
\end{equation}
This estimate is based on the assumption that the probability of correct assignment to a certain bin of a certain pair distance value is directly proportional to the difference $x$ between that value and the nearest bin boundary, or that $P_{ca}(x,dR) = a + b x$. Using the above assumed conditions for the values of $P_{ca}$ at the bin middle and on the boundary, we find that
\begin{eqnarray}
P_{ca}(x,dR)&=& (1+x/dR)/2 \mbox{ if } x < dR
\nonumber \\
P_{ca}(x,dR)&=& 1 \mbox{ if } x \geq dR 
\end{eqnarray}
Therefore, for the average probability of correct assignment we find that
\begin{eqnarray}
P(L,dR)&=& 1 - \frac{dR}{2L}  \mbox{ if } L \geq 2\ dR \mbox{  and }
\nonumber \\
P(L,dR)&=& 0.5+\frac{L}{8\ dR} \mbox{ if } L < 2\ dR.
\label{eq:Pca_liniar}
\end{eqnarray}
From this result we can infer that for $dR=0.4$\AA, if we choose a radial bin size $L=0.5$\AA, the confidence of correct bin assignment is about 65.6\%. In this work we employ a radial bin width of $L=1.2$\AA. This bin width is at least twice larger than the values employed by other authors. However, it ensures that we have a confidence level of at least 83.3\% that the radial bin assignment is correct.\footnote{If a Gaussian probability of correct assignment, $P_{ca}(x,dR)$, is used instead of the linear dependence assumed  above, with the same constraints ($P_{ca}(0,dR)=0.5$ and $P_{ca}(dR,dR)=1$) for the same value $dR=0.4$\AA, the confidence level for a bin size $L=1.2$\AA\  is estimated to be 87.3\%, and only 71.5\% for $L=0.5$\AA.} 

Similar considerations are used for studying the angular orientation pair distributions. Consider the polar coordinate system and the division of the $\theta$ and $\phi$ values. We assume that we have $N$ bins in $\theta$ and $2N$ bins in $\phi$. The solid angle corresponding to one angular bin is $d \Omega=2\pi/N^2$. This solid angle is proportional to the ratio between the area element that insures a certain high value for the confidence level of correct assignment, and the square of the corresponding minimum pair distance (radius) at which we expect that confidence level. Let us require a confidence level of 80\% at a distance of 5\AA. At larger distances the confidence level will be higher, but it will decrease at closer ranges. Since $d \Omega\approx L^2/(R+dR)^2$, and $L=1.2$\AA, $R=5.0$\AA\ and $dR=0.4$\AA, we have $d\Omega\approx 0.22^2=2\pi/N^2$. This reasoning leads us to employ the nearest even value of $N=12$ for the number of bins that divide the angular interval for $\theta$. Previous studies \cite{Jernigan-96} used a smaller value ($N=3$).

	The use of a larger radial bin size (1.2\AA\ vs. 0.5\AA) than previously suggested, is needed to ensure a high confidence level (at least 83.3\%) of correct radial bin assignment. Here, we employ twenty radial distance bins with $L=1.2$\AA\ for distances starting at $2$\AA. When we collect statistical data on the relative 3-dimensional orientation of pairs of residues, we obtain a similarly high confidence level by using $N=12$ for $\theta$ and $N=24$ for $\phi$.

	All the calculations presented in this section are based on the assumption that all protein structures analyzed have a resolution of 2\AA\ or better. If proteins of different structural resolution are analyzed, the optimal bin size can be estimated accordingly. These arguments assure that optimal radial and angular bin sizes (i.e. values that are small enough to provide a good resolution, yet large enough to correspond to a high statistical confidence level) are employed.

\end{document}